\documentstyle[osa,manuscript,epsf]{revtex}  

\begin{document}                
\title{
Quantum-fluctuation-induced repelling interaction of
quantum string between walls}
\author{Yoshihiro Nishiyama}
\address{
Department of Physics, Faculty of Science,
Okayama University \\
Okayama 700-8530, Japan}
%
\maketitle
\begin{abstract}                
Quantum string, which was brought into discussion recently as a model
for the stripe phase in doped cuprates, is simulated by means of
the density-matrix-renormalization-group method.
String collides with adjacent neighbors, as it wonders, owing to 
quantum zero-point fluctuations.
The energy cost due to the collisions is our main concern.
Embedding a quantum string between rigid walls with separation $d$,
we found that for sufficiently large $d$,
collision-induced energy cost obeys the formula
$\sim \exp (- A d^\alpha)$ with $\alpha =0.808(1)$, and
string's mean fluctuation width grows logarithmically $\sim \log d $.
Those results are not understood in terms of conventional picture
that the string is `disordered,' and only the short-wave-length
fluctuations contribute to collisions.
Rather, our results support a recent proposal that
owing to collisions,
short-wave-length fluctuations are suppressed,
but instead, long-wave-length fluctuations become significant. 
This mechanism would be responsible for stabilizing the stripe phase.
\end{abstract}
\section{Introduction}
\label{section1}

Recently, Zaanen brought up a problem of `quantum string'
\cite{Zaanen00},
which is
a line-like object subjected to line tension,
and it wonders owing to quantum-mechanical zero-point fluctuations.
Central concern is to estimate the interaction between adjacent strings
as it wonders quantum-mechanically and
undergoes entropy-reducing `collisions' with their neighbors.
Statistical mechanics of quantum-string gas would be relevant to 
the low-energy physics of
the so-called stripe phase observed experimentally in doped cuprates
\cite{Tranquada95,Wakimoto99,Noda99}.
In particular, one is motivated to gain insights how
the stripe pattern formed in cuprates acquires stability.
Actually, a good deal of theoretical analyses
had predicted tendency toward stripe-pattern formation 
\cite{Zaanen89,Schulz90,Kato90,Poilblanc89,Ichioka99,Kivelson98,Nayak97}.
However, first-principle simulations on the $t$-$J$ model
still remain controversial about that issue
\cite{Prelovsek93,White98,Hellberg99}.
The aim of the aforementioned recent analysis \cite{Zaanen00} is
to shed light, particularly, on the role of the
quantum fluctuations and the entropy-reducing collisions
at the expense of disregarding microscopic ingredients
of quantum string.

In the path-integral space-time picture, quantum string spans a
world sheet as time evolves \cite{Fradkin83}.
Hence, one may wonder that physics of quantum string might bear resemblance
to that of membrane, and so quantum string is readily understood
through resorting to the past outcomes about membrane 
\cite{Janke86,Janke87,Leibler89,Netz95,Gouliaev98,Kleinert99,Yamamoto00}.
However, nature of elasticities is different:
Quantum-string's
elastic energy is quadratic in its slope (line tension),
whereas membrane's is quadratic in its curvature (bending elasticity).
This seemingly slight difference causes, according to Ref. \onlinecite{Zaanen00},
quite distinctive behaviors.

When quantum string is laid down in a free space, its Hamiltonian
is
quadratic, and thus no peculiarity emerges.
(We will introduce the Hamiltonian afterwards.)
Yet, when it is squeezed by adjacent neighbors or walls,
there immediately arise awful complications due to many-body correlations.
In order to tackle with the string gas,
Zaanen employed Helfrich's technique \cite{Helfrich78,Helfrich84}, 
that has been utilized
in the course of the studies of stacked membranes.
Thereby, he arrived at the conclusion that
collision-induced energy cost $f$
would be described by
$f \sim \exp(-A d^{2/3})$ with string interval $d$.
That result cannot be understood in terms of conventional argument
\cite{Fisher82}
which gives $f\sim \exp(-A d^2)$; this argument is based on such picture
that string is `disordered,' and the collisions
are mainly due to
short-wave-length meandering modes.
(We will outline this argument in Sec. \ref{section3_2}.)
On the contrary,
Zaanen emphasizes the role of long-wave-length fluctuations,
because his theory is sensitive to 
infrared cutoff.

The purpose of this paper is to judge the validity of those scenarios by performing
first-principle simulations.
We put a quantum string between
rigid walls with spacing $d$, and measured its repelling interaction
by observing pressure against the walls.
(This trick has been used in the studies of the fluctuation pressure
of (classical) membrane \cite{Janke86,Janke87,Leibler89}.) 
The Hamiltonian, which we had simulated, is given by,
\begin{equation}
\label{Hamiltonian}
{\cal H}=\sum_{i=1}^{L} \left( \frac{p_i^2}{2m}
                                 + V(x_i) \right)
          +\sum_{i=1}^{L-1} \frac{\Sigma}{2} (x_i-x_{i+1})^2    . 
\end{equation}
Here, $x_i$ denotes the operator of transverse displacement of
a particle at $i$th site, and $p_i$ is its conjugate momentum. 
They satisfy the canonical commutation relations
$[x_i,p_j]={\rm i}\hbar \delta_{ij}$,
$[x_i,x_j]=0$ and
$[p_i,p_j]=0$.
$V(x)$ is rigid-wall potential with spacing $d$;
\begin{equation}
V(x)=
\left\{
\begin{array}{ll}
  0    &      {\rm for}   \    0 \le x \le d \\
\infty &      {\rm otherwise}
\end{array}
\right. .
\end{equation}
$\Sigma$ denotes line tension which puts particles into line. 
Classical version of this Hamiltonian has been used as a model
for line dislocations and steps on (vicinal) surfaces
\cite{Coppersmith82,Zaanen96}. 
Note that for sufficiently large $\Sigma$,
one can take continuum limit, with which one arrives at field-theoretical
version of quantum string.
Such continuum-limit version was studied analytically in Ref. 
\onlinecite{Zaanen00}.
For small $\Sigma$, however, continuum-limit treatment fails, and
our lattice-version Hamiltonian exhibits characteristics
which are not considered in the previous study.

The rest of this paper is organized as follows.
In the next section, we explicate our simulation scheme.
We used diagonalization method:
Note that elastic models such as ours (\ref{Hamiltonian}) have vast number of vibration
modes, which overwhelm computer-memory size.
Emphasis is laid upon the point how we had adopted
the idea of density-matrix renormalization group
\cite{White92,White93} to quantum string
so as to reduce the number of Hilbert-space bases.
Our algorithm is indebted to recent developments \cite{Zhang98}, where a polaron model
(lattice vibration)
was simulated with use of the density-matrix renormalization group; see also 
Ref. \onlinecite{Nishiyama99}.
Following the preparations,
in Sec. \ref{section3}, we perform numerical calculation.
We show that our first-principle simulation supports aforementioned Zaanen's scenario.
In the last section, we give summary and discussions.

\section{Details of numerical method: density-matrix renormalization group}
\label{section2}

In this section, we explain our simulation algorithm.
Emphasis is laid on the point how we had adopted the idea of the 
density-matrix renormalization group \cite{White92,White93} 
in order to treat quantum string.
Afterwards, we will show preliminary simulation data
so as to demonstrate its performance.

Quantum string is made of many particles
connected with line tension $\Sigma$.
Each particle spans infinite-dimensional Hilbert space.
Hence, one is forced to truncate, somehow, the number of bases
in order to diagonalize the Hamiltonian.
Even though one truncated bases of each particle,
the total bases of quantum string, as a whole, would exceed available
computer-memory size.
This difficulty arises inevitably in treating elastic (bosonic) degrees of freedom
by means of diagonalization. 
Recently, however, Zhang {\it et al.} reported that the difficulty is
overcome by an ingenious application of the density-matrix renormalization group
\cite{Zhang98}.
Actually, they
demonstrated that
a polaron model is diagonalized successfully;
see also Ref. \onlinecite{Nishiyama99}.
In the following, based on these recent developments, we propose
a scheme to diagonalize the quantum-string Hamiltonian (\ref{Hamiltonian}).

Our simulation algorithm proceeds recursively; after
the completion of one recursion (renormalization),
the number of treated particles (length of quantum string) 
is enlarged by two, and subsequent renormalization follows.
In the following, we explain
steps constituting one renormalization procedure:
Suppose that our system (quantum string) is decomposed into four parts,
that is, block, site, site and block, in this order.
After one renormalization is completed, composite systems of block and site 
are `renormalized'
into a new block with block Hilbert-space dimension {\em unchanged}.

As would be guessed,
`site' merely stands for one particle of quantum string.
We then need to prepare bases to represent the Hilbert space of `site.'
We had chosen such $M$ bases that are the 
eigenstates $|n\rangle_{\bullet}$ ($n=1 \sim M$)
of the intra-site Hamiltonian
${\cal H}_{\bullet}=p^2/(2m)+V(x)$ with energy $E_n = \hbar^2 \pi^2 n^2/(2m d^2)$.
(Therefore,
the direct product of $|n\rangle_{\bullet}$
gives the eigenstate of the total Hamiltonian ${\cal H}$,
provided that the line tension is tuned off ($\Sigma=0$).)
With respect to these bases $\{ |n\rangle_{\bullet} \}$ ($n=1$-$M$).
we represent 
the matrix of the $x_i$ operator;
\begin{equation}
\label{x_dot}
[x_\bullet]_{nm}=
{}_\bullet \langle n | x | m \rangle_\bullet = 
\left\{
\begin{array}{ll}
\frac{2d}{\pi^2}\left(   \frac{1}{(m+n)^2}-\frac{1}{(m-n)^2}
                   \right)  & (n,m)={\rm (even,odd)\ or\ (odd,even)}
                                                                      \\
0 & {\rm otherwise}
\end{array}
\right.   .
\end{equation}
The truncation bound $M$ is one source of numerical errors.
We need to choose $M$ sufficiently large;
we will demonstrate afterwards that this truncation does not
deteriorate simulation precision in practice.

Let us turn to explaining `block.'
The `block' stands for a part (fragment) of quantum string, and 
actually, it contains many particles in it.
Hence, at a glance, one may wonder that the Hilbert-space dimension of block
would be extremely large.
Yet, owing to the density-matrix renormalization, the dimension is reduced
so that it can be stored in computer memory.
The Hilbert space of block is spanned by those
bases $|n\rangle_{\rm B}$
($n=1 \sim m$).
The bases are to be prepared in the {\em preceding} renormalization procedure;
see below.
With respect to these bases, 
one has to represent
the intra-block Hamiltonian ${\cal H}_{\rm B}$ and
the coordinate operator
of the particle at the end of `block'
$x_{\rm B}$; see below.
At
the initial stage of renormalization, 
the block is merely a `site,'
and so we start with  ${\cal H}_{\rm B}={\cal H}_{\bullet}$,
$x_{{\rm B}}=x_\bullet$ and $m=M$.

Provided that the above matrices are at hand,
the (total) Hamiltonian of quantum string is expressed in terms of them;
\begin{eqnarray}
{\cal H} =  {\cal H}_{{\rm B}\bullet\bullet{\rm B}} &=&
 {\cal H}_{\rm B} \otimes \hat{1} \otimes \hat{1} \otimes \hat{1}
 +\hat{1} \otimes {\cal H}_\bullet \otimes \hat{1} \otimes \hat{1}  \nonumber \\
& & +\hat{1} \otimes \hat{1} \otimes {\cal H}_\bullet \otimes \hat{1}
    +\hat{1} \otimes \hat{1} \otimes \hat{1} \otimes {\cal H}_{\rm B}  \nonumber \\
& & +\frac{\Sigma}{2}(x_{\rm B}^2 \otimes \hat{1} \otimes \hat{1} \otimes \hat{1}
      -2 x_{\rm B} \otimes x_\bullet \otimes \hat{1} \otimes \hat{1}
      +\hat{1} \otimes x_\bullet^2 \otimes \hat{1} \otimes \hat{1} )  \nonumber \\
& & +\frac{\Sigma}{2}(\hat{1} \otimes x_\bullet^2 \otimes \hat{1} \otimes \hat{1}
      -2 \hat{1} \otimes x_\bullet \otimes x_\bullet \otimes \hat{1} 
      +\hat{1} \otimes \hat{1} \otimes x_\bullet^2 \otimes \hat{1} )  \nonumber \\
& & +\frac{\Sigma}{2}(\hat{1} \otimes \hat{1} \otimes x_\bullet^2 \otimes \hat{1}
      -2 \hat{1} \otimes \hat{1} \otimes x_\bullet \otimes x_{\rm B} 
      +\hat{1} \otimes \hat{1} \otimes \hat{1} \otimes x_{\rm B}^2)  .
\end{eqnarray} 
Diagonalizing this matrix, one obtains 
the ground state,
\begin{equation}
|\psi\rangle=\sum_{ijkl} \psi_{ijkl} 
  |i\rangle_{\rm B}  |j\rangle_\bullet  |k\rangle_\bullet 
      |l\rangle_{\rm B}  .
\end{equation}
With use of $\psi_{ijkl}$, we obtain the density matrix for the left-half
sub-system (${\rm B}+\bullet$);
\begin{equation}
[\rho]_{i,j;i',j'}
     = \sum_{kl} \psi_{ijkl}\psi^*_{i'j'kl} .
\end{equation}
Diagonalizing this, we obtain the eigenstates and the eigenvectors;
$\rho |u_n\rangle = w_n |u_n\rangle$ 
with $w_1>w_2>\cdots>w_{(M \times m)}$.
Those bases $| u_n \rangle$ with large weight $w_n$ would be important (relevant)
in order to describe the physics of the sub-system of block and site.
Therefore, we store the bases $| u_n\rangle$ with $n=1$-$m$, 
and discard the others.
This criterion is the essence of the so-called density-matrix renormalization
group \cite{White92,White93}.
That truncation may cause another source of numerical errors.
Later,
we will demonstrate that this error is very small.
Finally, we perform the `density-matrix renormalization';
\begin{eqnarray}
[{\cal H}_{{\rm B}'}]_{nm} &=&
\left\langle u_n \left|
    {\cal H}_{\rm B}\otimes \hat{1}  + \hat{1} \otimes {\cal H}_\bullet
  + \frac{\Sigma}{2} \left(  x_{\rm B}^2 \otimes \hat{1} -2 x_{\rm B} \otimes x_\bullet 
            + \hat{1} \otimes x_\bullet^2   \right)
\right| u_m \right\rangle          \\
 \left[ x_{{\rm B}'} \right]_{nm} 
  &=&
\left\langle u_n 
\left|
\hat{1} \otimes x_\bullet
\right| 
u_m \right\rangle                         .
\end{eqnarray}
Now, a renormalization is completed.
We can restart the next renormalization from the beginning, replacing
the renormalized block ${\rm B}'$ with ${\rm B}$.
It is to be noted that
through the renormalization,
the block dimension is kept within $m$.

\begin{figure}
\begin{center}\leavevmode
\epsfbox{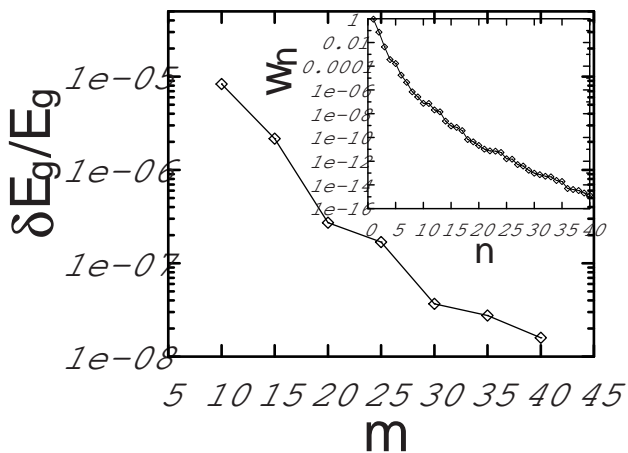}
\end{center}
\caption{
Relative error of the ground-state energy for the system
with $d=4$, $\Sigma=4$, $L=8$ and $M=9$.
`Exact' energy is calculated with respect to the full Hilbert-space dimensions 
$M^L=9^8$,
while `approximate' energy is calculated for truncated `block'
bases $m$ with use of the density-matrix renormalization group.
Ground-state-energy difference between them gives $\delta E_{\rm g}$.
Note that truncated-base calculation with small $m$ 
reproduces full-diagonalization result very precisely.
Inset shows the density-matrix eigenvalues $\{ w_n \}$, which are
used for monitoring simulation precision;
see text for details.}
\label{fig_relative_error}
\end{figure}

In the following, we will show that the above algorithm actually works.
In Fig. \ref{fig_relative_error}, we plotted the relative error of the ground-state energy
$\delta E_{\rm g}/E_{\rm g}$
for the system with $\Sigma=4$ and $d=4$.
(This parameter condition is of great physical significance as would be
shown in Sec. \ref{section3}.)
The system size is $L=8$, and the dimension of `site' is $M=9$.
Therefore, the full number of bases is $M^L = 43046721$,
 which
is about to exceed the limit of available computer-memory size
and is barely manageable with full diagonalization method.
For these full bases, we calculated the `exact' ground-state energy,
while 
`approximate' energy is calculated by means of the density-matrix renormalization
group with truncated `block' dimension $m$.
The energy difference between them gives $\delta E_{\rm g}$.
The relative error $\delta E_{\rm g}/E_{\rm g}$ is plotted in Fig. \ref{fig_relative_error}.
We achieve very small error $10^{-8}$ with $m=20$,
for which the total number of bases is no more than $20^2 \cdot 8^2=25600$.
Hence, we see that our algorithm works efficiently.
The inset of Fig. \ref{fig_relative_error} shows the distribution of the density-matrix
eigenvalues $\{ w_n \}$.
Usually, $w_n$ is utilized for monitoring $\delta E_{\rm g}/E_{\rm g}$,
because they look alike.
However, in our case, there are discrepancies in their magnitudes.
The discrepancies may be, perhaps, due to the fact that our Hamiltonian-matrix elements 
distribute over wide range.  
By the way, we chose $m$ corresponding to
$w_m=10^{-12}$ in Sec. \ref{section3}.
In this way, we kept precision within $10^{-7}$.
Typically, we need, at most, 
$m=30$ bases.

\begin{figure}
\begin{center}\leavevmode
\epsfbox{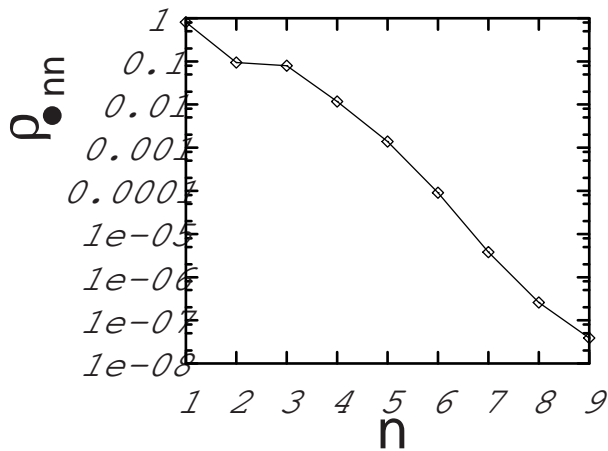}
\end{center}
\caption{
Probability weight $[\rho_\bullet]_{nn}$ of the state $|n\rangle_\bullet$ is plotted.
This plot indicates how site-dimension truncation $M$
affects the reliability of simulation.
Probability weight for $n=9$ is negligibly small; 
$[\rho_\bullet]_{99} \approx 10^{-7}$.
Hence, truncation of those states with $n>9$ does not deteriorate any reliability 
in practice.
In Sec. {\protect \ref{section3}}, we impose even severer 
request $[\rho_\bullet]_{MM} \approx 10^{-10}$.}
\label{fig_rho_dot}
\end{figure}

In the above, we have checked that the truncation of block dimension $m$ does not
harm any reliability of our simulation.
Finally, we will examine the influence of site-dimension truncation $M$.
In order to elucidate that, it is sensible to monitor, 
\begin{equation}
[ \rho_\bullet ]_{jj'} = \sum_{ikl} \psi_{ijkl}\psi^*_{ij'kl}  .
\end{equation}
As would be apparent from the definition, $[ \rho_\bullet ]_{nn}$ tells
the degree of significance of 
the state $| n \rangle_\bullet$.
Because we use the bases of $n=$1-$M$ and discard the others,
it should be checked whether $[\rho_\bullet]_{MM}$ is small enough.
We see from Fig. \ref{fig_rho_dot}, that the $M=9$ state is of
very rare probability $10^{-7}$.
In the subsequent simulations in Sec. \ref{section3}, we impose
even severer request $[\rho_\bullet]_{MM}=10^{-10}$.
In order to match this request, we need, at most, $M=20$ bases;
hence, the maximal total number of Hilbert-space bases is no more than
$m^2\cdot M^2=30^2 \cdot 20^2=360000$.
As is mentioned above, 
the full diagonalization of Figs.
\ref{fig_relative_error}-\ref{fig_rho_dot}
requires the Hilbert-space 
dimensions $43046721$ for string length $L=8$.
It is far beyond the
capability of the diagonalization method to treat longer string, unless we resort
to the density-matrix renormalization group.

\section{Numerical results and discussions}
\label{section3}

In this section, we present simulation results.
In our simulation, we treated sufficiently large system sizes by repeating
renormalizations, until simulation result converges to a (thermodynamic) limit.
System parameters of $m$ and $\hbar$ are fixed; 
namely, we set $m=1$ and $\hbar=1$.
those parameters just fix the coefficient of the kinetic-energy term.
Therefore,
the choice of parameters does not harm any generality.
Technical details are to be referred to the previous section.

\subsection{Collision-induced energy cost $f$ and elasticity modulus $B$}
\label{section3_1}

\begin{figure}
\begin{center}\leavevmode
\epsfbox{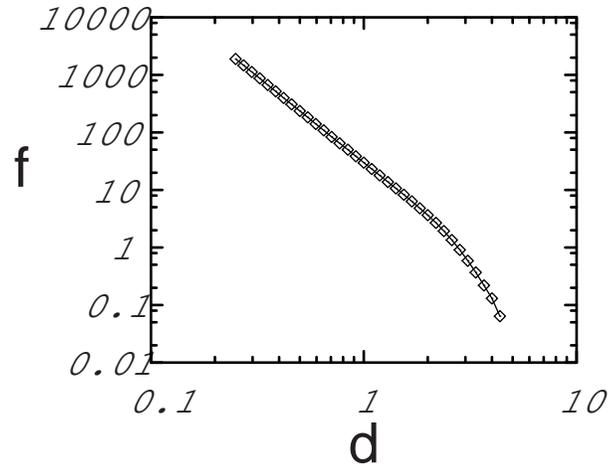}
\end{center}
\caption{
Excess energy cost due to collisions $f$ ({\protect \ref{collision_f}}) is plotted for 
$\Sigma=4$ and various $d$.
We see that for $d<2$, it obeys power law $f \sim d^{-3}$, while
for $d>2$, it drops rapidly.}
\label{fig_f}
\end{figure}

In Fig. \ref{fig_f}, we plotted the collision-induced energy cost,
\begin{equation}
\label{collision_f}
f=\frac{ E_{\rm g}(d)-E_{\rm g}(d\to\infty) }{ d } ,
\end{equation}
for the system with line tension $\Sigma=4$ and various wall spacing $d$.
Here, $E_{\rm g}(d)$ denotes the ground-state
energy per one particle for wall spacing $d$.
As would be apparent from the definition (\ref{collision_f}),
$f$ measures excess energy cost (per unit volume) due to the presence of walls.
It is notable that
$E_{\rm g}(d\to\infty)$ is calculated exactly,
\begin{equation}
E_{\rm g}(d\to\infty)= \frac{1}{2\pi} \int_{-\pi}^{\pi} {\rm d}k \sqrt{\frac{\Sigma}{m}}
    \left| \sin \frac{k}{2} \right|              ,
\end{equation}
because for $d\to\infty$,
the Hamiltonian reduces to quadratic form.

In Fig. \ref{fig_f}, we notice that two distinctive regimes exist:
For $d<2$,
the collision energy $f$
decreases obeying power law,
\begin{equation}
\label{f_small_d}
f \sim 1/d^3,
\end{equation}
whereas for $d>2$, it decays `rapidly';
actually, it is our main concern to clarify how `rapid' it decays.
The behavior for $d<2$ is understood immediately:
Suppose that the line-tension is turned off ($\Sigma=0$),
each particle becomes independent, and it
reduces to a text-book problem of `particle in a box,'
for which
the ground-state energy is solved exactly;
$E_{\rm g}=\hbar^2 \pi^2/(2m d^2)$.
Hence, we arrive at the relation $f=E_{\rm g}/d\sim 1/d^3$.
To summarize, for small $d<2$,
inter-particle interaction
is irrelevant.
On the other hand, for large $d$, the inter-particle interaction 
may become relevant.
Because of this,
the particles become correlated, and $f$ drops very rapidly.

In order to elucidate characteristics of fluctuation-induced interaction,
it is sensible to calculate
the elasticity modulus which is defined by the formula,
\begin{equation}
\label{modulus}
B= d^2 \frac{\partial^2 f}{\partial d^2}  .
\end{equation}
In Fig. \ref{fig_B},
we plotted $B$ for the same parameter range as that of Fig. \ref{fig_f}.
As well as $f$, $B$ also exhibits two distinctive regimes.
For $d<2$, we see that $B$ decays in the form $B\sim 1/d^3$.
This behavior is understood immediately, if we remember $f\sim 1/d^3$.
The behavior for large $d>2$ is still remaining mysterious.
We will explore this in the next subsection.

\subsection{Scaling analyses}
\label{section3_2}

\begin{figure}
\begin{center}\leavevmode
\epsfbox{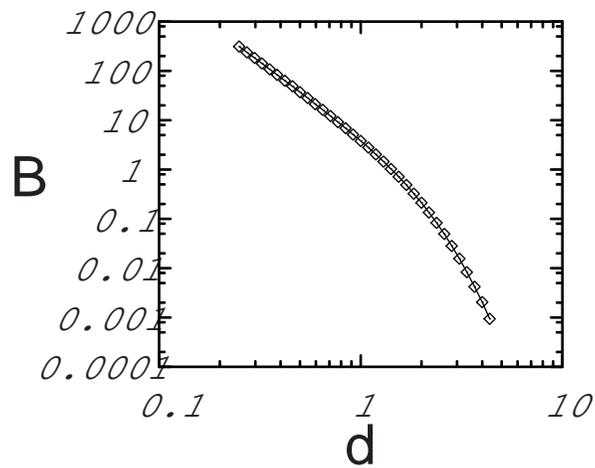}
\end{center}
\caption{
Elasticity modulus $B$ ({\protect \ref{modulus}}) is plotted for 
the same parameter range as that of Fig. {\protect \ref{fig_f}}.
We see that for $d<2$, it obeys power law $B \sim d^{-3}$, while
for $d>2$, 
$B$ drops rapidly.}
\label{fig_B}
\end{figure}

Here, we re-examine the data of $f$ and $B$ presented in Figs. \ref{fig_f} and
\ref{fig_B}.
For that purpose, we need to resort to past outcomes on $f$ and $B$.
First, we summarize recent remarkable predictions by Zaanen \cite{Zaanen00}.
In his theory, 
$B$ works as a mean field, which is to be determined self-consistently.
The self-consistency equation to be solved is,
\begin{equation}
\label{SC_condition}
f = C \frac{B}{\Sigma d^2} \left(
          \log \left( \frac{\Sigma d}{B} \right)
               + C' \right) .
\end{equation}
Appearance of logarithmic term brings about subtleties.
It is hard to solve this self-consistency equation.
However, 
asymptotic form for $d\to\infty$
is calculated as follows;
\begin{eqnarray}
\label{f_Zaanen}
f & \sim & {\rm e}^{- C'' d^{2/3}}            \\
B & \sim &  d^2 {\rm e}^{-C'' d^{2/3}}   .
\end{eqnarray}
In addition to Eq. (\ref{SC_condition}), 
under the assumption that the string-meandering modes
are `compactified' (frozen),
he found another
self-consistency condition,
\begin{equation}
\label{SC_condition2}
f = C''' \frac{ \sqrt{B} }{d^{3/2}}  .
\end{equation}
Note that the validity of this relation is checked, if we assume 
Eq. (\ref{f_small_d}).
Hence,
this relation (\ref{SC_condition}) may be realized for small $d$.

Besides those recent treatments,
there exists an ingenious argument
\cite{Fisher82} for tackling with entropic interaction.
The argument, applied for quantum string (\ref{Hamiltonian}), yields predictions 
different from the above.
Below, we outline this argument:
As is mentioned in Introduction, in the path-integral picture,
quantum string bears resemblance to
(classical) membrane.
When quantum string is embedded in a free space, 
we can solve
its mean deviation $\sim (\log l)^{1/2}$ with membrane's
linear dimension $l$.
Suppose that the wall width (mean membrane interval) is $d$,
with use of this relation, one obtains an estimate of the 
surface $S$ per one collision such as
$d \sim (\log \sqrt{S})^{1/2}$.
Assuming that each collision (contact-or-crossing)
gives rise to entropy loss $\sim k_{\rm B} \log 2$,
we obtain the collision-induced
energy gain per unit surface;
\begin{equation}
\label{f_Fisher}
\sim k_{\rm B} \log 2 {\rm e}^{- C'' d^2}.
\end{equation}
The argument is based on the assumption that
the string would be disordered spatially as in the Einstein-like view of a crystal.
In other words, short-wave-length fluctuations contribute to
collisions.
On the contrary, Ref. \onlinecite{Zaanen00} emphasizes the significance of
long-wave-length fluctuations.
Our aim is to judge which picture is valid by means of first-principle simulations.

\begin{figure}
\begin{center}\leavevmode
\epsfbox{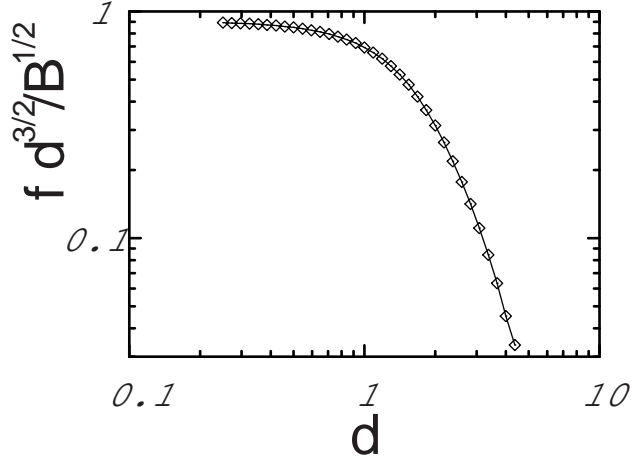}
\end{center}
\caption{
$f d^{3/2} / B^{1/2} $ is plotted with use of the data shown in
Figs. {\protect \ref{fig_f}} and {\protect \ref{fig_B}}.
This plot shows that the relation ({\protect \ref{SC_condition2}}),
which is valid for 
`compactified' string, is realized
for small $d$.
On the other hand, for $d>2$, it does not hold at all.
}
\label{fig_scaling1}
\end{figure}

Guided by the above ideas, we will carry out scaling analyses.
First, in Fig. \ref{fig_scaling1}, we plotted $f d^{3/2}/\sqrt{B}$ 
with use of the data shown in Figs. \ref{fig_f} and \ref{fig_B}.
Note that $f d^{3/2} / \sqrt{B}$ should be constant, if 
Eq. (\ref{SC_condition2}) holds.
The scaling plot indicates that
for $d<2$, in fact, the scaled data are kept constant.
On the contrary, for $d>2$, the scaled data drop suddenly,
suggesting that Eq. (\ref{SC_condition2}) does not hold any more.
In consequence, we confirmed that for small $d$,
string-meandering modes are `compactified' (irrelevant).
These observations are consistent with those found in the previous subsection.

\begin{figure}
\begin{center}\leavevmode
\epsfbox{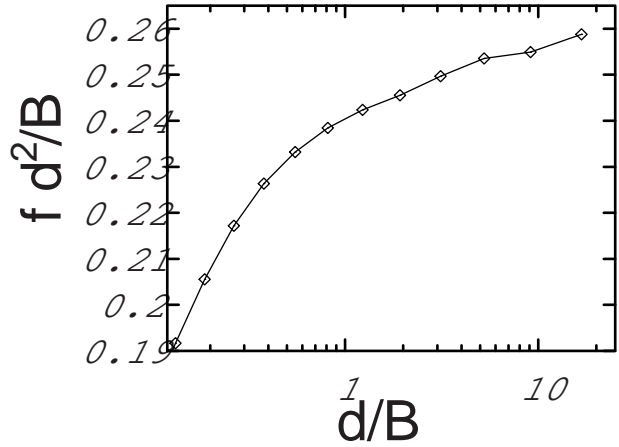}
\end{center}
\caption{
$f d^2 / B$ is plotted against $d/B$ with use of the data shown in
Figs. {\protect \ref{fig_f}} and {\protect \ref{fig_B}}.
For $d/B>1$, the data approach to a straight line asymptotically.
This result indicates that
the relation ({\protect \ref{SC_condition}}) is satisfied for the
scaling region $d/B>1$.
}
\label{fig_scaling2}
\end{figure}

Secondly, let us turn to large-$d$ regime.
In Fig, \ref{fig_scaling2},
we plotted $f d^2 / B$ against $d/B$ with use of the data
shown in Figs. \ref{fig_f} and \ref{fig_B}.
Note that the scale of abscissa is logarithmic.
Therefore,
the data should align, if the relation (\ref{SC_condition}) is satisfied.
In fact, 
for scaling regime $d/B>1$, we see that the scaled data get aligned,
and exhibit a positive slope.
This result justifies the validity of Eq. (\ref{SC_condition}).
Namely, the mean-field treatment of Ref. \onlinecite{Zaanen00} appears to
capture
the essence of this physics.
For exceedingly large $d$, eventually, numerical data become scattered (unstable)
gradually.
This is a symptom of numerical errors.
Note that for exceedingly large $d$, the elasticity modulus $B$ is very small
so that it suffers from tiny numerical errors.
Hence, we cannot continue simulations for extremely large $d$.

\begin{figure}
\begin{center}\leavevmode
\epsfbox{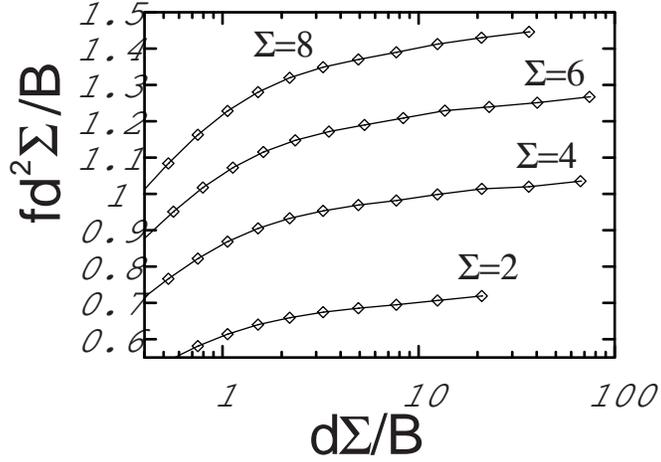}
\end{center}
\caption{
Similar scaling analysis as Fig. {\protect \ref{fig_scaling2}} but for
various $\Sigma=2$-$8$.
In the scaling regime $d\Sigma/B>3$,
slopes appear to be identical, suggesting that
the constant $C$ in Eq. ({\protect \ref{SC_condition}}) is universal 
with respect to arbitrary $\Sigma$.
}
\label{fig_scalings}
\end{figure}

In Fig. \ref{fig_scalings}, we presented similar scaling data but for various 
$\Sigma=2$, $4$, $6$ and $8$.
The plots show that the scaling-plot slopes are identical.
This fact tells that the constant $C$ in Eq. (\ref{SC_condition}) is indeed 
universal with respect to $\Sigma$.
It is, however, suggested that $C'$ is subjected to a correction to scaling,
because plots do not overlap precisely.
We had found that
scaling data for
very small $\Sigma<1$ are not described by Eq. (\ref{SC_condition}).
As a matter of fact, the slopes are almost vanishing.
Breakdown of Eq. (\ref{SC_condition}) for very small $\Sigma$
 is reasonable, because the equation
is derived under continuum-limit treatment
which is not justified for very small $\Sigma$.
Therefore, such the region of small $\Sigma$ lies out of 
the scope of Eq. (\ref{SC_condition}).
By the way, we stress that
our first-principle simulation does cover any conditions.

\subsection{Asymptotic form of $f$}
\label{section3_3}

\begin{figure}
\begin{center}\leavevmode
\epsfbox{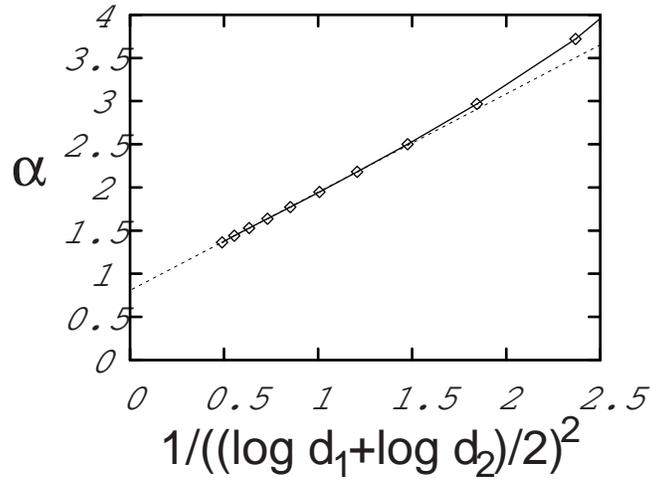}
\end{center}
\caption{
Exponent $\alpha$ ({\protect \ref{exponent}}) is plotted 
with use of the data shown in Fig. {\protect \ref{fig_f}}.
With the least-square method, we extrapolated the data
so as to obtain
$\alpha=0.808(1)$ for $d\to\infty$. 
}
\label{fig_exponent}
\end{figure}

In order to confirm the above observation, we investigate
the asymptotic form of $f$ for $d\to\infty$.
As is mentioned in the previous subsection, 
$f$ should obey the asymptotic form,
\begin{equation}
f \sim {\rm e}^{- C'' d^\alpha}                    ,
\end{equation}
with the exponent either $\alpha=2/3$ (\ref{f_Zaanen}) or $\alpha=2$ (\ref{f_Fisher}).
We expect that the former would be realized,
because it is derived from Eq. (\ref{SC_condition}), whose validity is checked in 
Sec. \ref{section3_2}. 
We calculated
$\alpha$ by means of the formula,
\begin{equation}
\label{exponent}
\alpha=
  \frac{ \log \left( \log f(d_1) / \log f(d_2) \right) }{ \log(d_1/d_2) }  ,
\end{equation}
with respect to adjacent two data points of $d=d_1$ and $d_2$ depicted 
in Fig. \ref{fig_f}.
We plotted $\alpha$ in
Fig. \ref{fig_exponent}.
The scale of abscissa 
$1/((\log d_1 + \log d_2 )/2)^2$ 
is chosen so as to achieve straight data alignment.
We employed the least-square method in order to extrapolate the result for
$d\to\infty$, and thereby,
we obtained the estimate
$\alpha=0.808(1)$.
This result clearly supports
$\alpha=2/3$ (\ref{f_Zaanen}) rather than
$\alpha=2$ (\ref{f_Fisher}).
We notice that the convergence speed is rather slow;
note that the abscissa scale of Fig. \ref{fig_exponent} is logarithmic.
Hence, we found that the asymptotic form (\ref{f_Zaanen}) is realized
for extremely large $d$.

\subsection{Mean fluctuation width $\Delta$}
\label{section3_4}

\begin{figure}
\begin{center}\leavevmode
\epsfbox{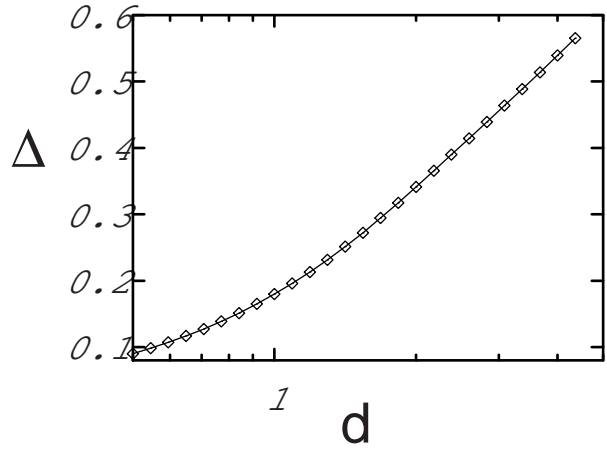}
\end{center}
\caption{
Meandering-fluctuation width ({\protect \ref{deviation}}) is 
plotted for the same parameter range
as that of Fig. {\protect \ref{fig_f}}.
For $d>2$, $\Delta$ grows logarithmically ($\Delta \sim \log d$).
That is, string's fluctuation is far less than the wall spacing $d$.
This result indicates that the string is straightened by collisions.
}
\label{fig_Delta}
\end{figure}

So far, we have confirmed the validity of the relations (\ref{SC_condition})
and (\ref{f_Zaanen}).
Underlying physics of these relations are rather astonishing \cite{Zaanen00}:
It is speculated that collisions rather contribute to straightening the quantum string.
In this subsection, we will examine this remarkable scenario through observing
the
mean fluctuation deviation,
\begin{equation}
\label{deviation}
\Delta = \sqrt{ \langle x_i^2 \rangle -
   \langle x_i  \rangle^2 } .
\end{equation}
In Fig. \ref{fig_Delta},
we plotted $\Delta$
for the same parameter conditions as in Fig. \ref{fig_f}.
To our surprise, $\Delta$ grows logarithmically ($\Delta \sim \log d$) for large $d$.
That is,
$\Delta$ is much less than the wall spacing $d$.
This feature contradicts our intuition 
that string would be short-range disordered, and the collisions would be
due to microscopic fluctuations.
Because Eq. (\ref{f_Fisher}) relies on this picture, it turned out consequently, that
the picture is incorrect for quantum string.
Rather, our result 
supports the aforementioned speculation that string acquires macroscopic stiffness.

\section{Summary} 
\label{section4}

We have investigated fluctuation-induced repelling interaction of quantum string
described by the Hamiltonian (\ref{Hamiltonian}).
First, we have developed simulation scheme based on the idea
of the density-matrix renormalization group.
We found that the scheme works very efficiently.
As is demonstrated in Fig. \ref{fig_relative_error}, ground-state energy is precise up to
the eighth digit.
Precision is crucial in our study, because we need 
to calculate second-order derivative in order to obtain
the elasticity modulus
$B$
(\ref{modulus}).
Secondly, based on those preparations, we have performed extensive simulations.
Simulation data suggest that two distinctive regimes exist:
For small $d$, intra-particle interaction dominates so that
simulation data are understood by ignoring line tension
($\Sigma=0$).
That is, collision-induced energy cost $f$ and the elasticity modulus
$B$ obey simple formulae such as $f,B\sim d^{-3}$.
For large $d$, on the other hand, 
particles get correlated by
line tension $\Sigma$, and physics becomes much harder to interpret.
We made trials of
several types of scalings in Figs. \ref{fig_scaling1}-\ref{fig_scalings}.
From those scaling plots, in consequence, we found that the data of $f$ and $B$ agree with
Zaanen's self-consistency condition (\ref{SC_condition}).
Moreover, we 
investigated the asymptotic form of $f$.
We found $f \sim \exp(- C' d^{\alpha})$ with $\alpha=0.808(1)$,
which again supports Zaanen's result (\ref{f_Zaanen}) rather than Eq. (\ref{f_Fisher}).

According to him, underlying physics of Eqs. (\ref{SC_condition}) 
and (\ref{f_Zaanen}) is quite peculiar;
collisions with adjacent neighbors rather suppress short-range fluctuations,
and consequently, quantum string is straightened macroscopically.
In fact, we found that in Fig. \ref{fig_Delta}, 
string's 
fluctuation deviation 
$\Delta$ is bounded within $\Delta \sim \log d$, which is far less than the
wall spacing $d$.
The above observations tell that in essence, quantum-string's ground state
is governed by infrared fluctuations.
In this respect, it is decisively important to treat long quantum string;
otherwise we could not have observed those features mentioned above.
Hence, our simulation scheme is advantageous.

We have confirmed that
owing to collisions, actually,
order out of disorder sets in.
This observation immediately leads an expectation that the stripe pattern
observed in doped cuprates
\cite{Tranquada95,Wakimoto99,Noda99}
is stabilized by this mechanism.
To verify this definitely,
one may need to explore `stacked' strings.
As for stacked membranes, it has been known that
$N$-membrane physics is essentially the same as that of a single
membrane confined between walls \cite{Janke87,Kleinert99}.
That is, there is a nice relation, with which one readily obtains $N$-membrane 
behavior with use of {\em single}-membrane result.
It remains for future study to verify that similar relation holds
as well as for stacked strings.

\section*{Acknowledgment}
Numerical calculation was performed on Alpha workstations
of theoretical physics group, Okayama university.

\end{document}